\documentclass[11pt,a4paper,oneside]{article}
\usepackage[top=3cm, bottom=3cm, left=2cm, right=2cm]{geometry}
\usepackage[english]{babel}
\usepackage[utf8x]{inputenc}
\usepackage[T1]{fontenc}

\usepackage{amsthm,amsmath,amssymb,mathrsfs,dsfont,mathtools}
\usepackage{microtype}

\usepackage{bm}
\usepackage{graphicx}
\usepackage[colorinlistoftodos]{todonotes}
\usepackage[colorlinks=true, allcolors=blue]{hyperref}
\usepackage{booktabs,subcaption}
\usepackage{hyperref}
\usepackage[all]{nowidow}

\usepackage{setspace}
\usepackage[ruled,norelsize,vlined]{algorithm2e}
\usepackage{tikz} % For the plot

\usepackage[sf]{titlesec}
\usepackage{sectsty}
\allsectionsfont{\centering \normalfont\scshape} % Make all sections centered, the default font and small caps

\def\T{{ \mathrm{\scriptscriptstyle T} }}

\usepackage[round]{natbib}

\usepackage{authblk}
\title{Conjugate priors and bias reduction for logistic regression models}
\date{}
\author[1]{Tommaso Rigon}
\author[2]{Emanuele Aliverti}
\affil[1]{Department of Economics, Management and Statistics, University of Milano--Bicocca, 20126 Milano, Italy}
\affil[2]{Department of Economics, University Ca' Foscari, 30121, Venezia, Italy}

\newtheorem{theorem}{Theorem}

\theoremstyle{definition}

\makeatletter
\def\env@cases{%
  \let\@ifnextchar\new@ifnextchar
  \left\lbrace
  \def\arraystretch{0.8}%
  \array{@{}l@{\quad}l@{}}}
\makeatother

\begin{document}
\maketitle

\begin{abstract}
	Logistic regression models for binomial responses are routinely used in statistical practice. However, the maximum likelihood estimate may not exist due to data separability. We address this issue by considering a conjugate prior penalty which always produces finite estimates. Such a specification has a clear Bayesian interpretation and enjoys several invariance properties, making it an appealing prior choice. We show that the proposed method leads to an accurate approximation of the reduced-bias approach of \citet{Firth1993}, resulting in estimators with smaller asymptotic bias than the maximum-likelihood and whose existence is always guaranteed. Moreover, the considered penalized likelihood can be expressed as a genuine likelihood, in which the original data are replaced with a collection of pseudo-counts.  Hence, our approach may leverage well established and scalable algorithms for logistic regression. We compare our estimator with alternative reduced-bias methods, vastly improving their computational performance and achieving appealing inferential results. 
\end{abstract}

\section{Introduction}\label{sec:intro}

Logistic regression is arguably one of the most widely used generalized linear models in statistical practice. In such a model, it is assumed that each entry of the vector $y = (y_1, \dots, y_n)^\T$ is a realization of independent binomial random variables with number of trials $m_1, \dots, m_n$ and success probabilities $\pi_1,\dots, \pi_n$. Moreover, suppose that each response $y_i$ is associated with a $p$-dimensional covariate vector $x_i = (x_{i1}, \dots, x_{ip})^\T$, for $i=1,\dots,n$. Then, a logistic regression model has
\begin{equation}
	\label{eq:mod_log}
	(y_i \mid m_i, \pi_i) \sim \mbox{Bin}(m_i,\pi_i), \qquad \pi_i = \frac{\exp{(x_i^\T\beta) }}{1 + \exp{(x_i^\T\beta) }}, \qquad (i=1,\dots,n),
\end{equation}
where $\beta = (\beta_1, \dots, \beta_p)^\T$ is a $p$-dimensional vector of unknown regression coefficients. We customarily assume that the $n \times p$ design matrix $X$, whose rows are $x_1,\dots,x_n$, is of full rank. Thus, the log-likelihood function is
\begin{equation}\label{eq:loglik}
\ell(\beta; y) = \sum_{i=1}^n y_i (x_i^\T\beta) - \sum_{i=1}^n m_i \log\{1 + \exp(x_i^\T\beta)\},
\end{equation}
and the maximum likelihood estimate $\hat{\beta}$ of $\beta$ is the maximizer of~\eqref{eq:loglik}. Unfortunately, the maximum likelihood estimate may not exist, meaning that at least one component of the vector $\hat{\beta}$ is infinite.  It is well known that such an unpleasant phenomenon occurs in presence of data separation, as carefully described in \citet{Albert1984}. If separation occurs, standard procedures % such as the iteratively reweighted least squares algorithm \citep{Green1984} 
may fail to converge, resulting in degenerate predicted success probabilities and misleading inferential conclusions.

In a seminal paper, \citet{Firth1993} showed that the maximizer of a suitable penalized likelihood, obtained by considering a Jeffrey's prior penalty, has a smaller asymptotic bias compared to the maximum likelihood estimator, while \citet{Kosmidis2021} proved that Firth's bias-reduced estimates always exist and solve the separability issue.
The finiteness and shrinkage properties of Firth's reduced-biased estimates spurred an interesting line of research, and specialized algorithms are described in \citet{Kosmidis2010, Kosmidis2020}. Penalized methods for logistic regression models have become increasingly popular in health and medical sciences, and numerous developments of the original approach of \citet{Firth1993} have been proposed in such specialized literature \citep[e.g.,][]{Greenland2015,Puhr2017}.

In high-dimensional settings, that is when $p$ and $n$ are both large, existing implementations of \citet{Firth1993} may be computationally too demanding;  refer  for instance to \citet{Sur2019}. Instead, we propose a bias-reduction approach which retains Firth's appealing properties but it is much easier to implement, since it relies on a simple perturbation of the data. Specifically, we penalize the logistic regression likelihood by the conjugate prior of \citet{Diaconis1979}, resulting in the following penalized log-likelihood
\begin{equation}\label{eq:penalized}
\tilde{\ell}(\beta; y) =  \ell(\beta; y) + \frac{p}{2m} \sum_{i=1}^n m_i (x_i^\T\beta) - \frac{p}{m} \sum_{i=1}^n m_i \log\{1 + \exp(x_i^\T\beta)\},
\end{equation}
where $m = \sum_{i=1}^n m_i$. The penalized log-likelihood~\eqref{eq:penalized} can be rewritten, up to a multiplicative constant, in terms of a genuine log-likelihood, having replaced the actual data $y$ with a vector of pseudo-counts $\tilde{y} = (\tilde{y}_1,\dots,\tilde{y}_n)^\T$, defined as
\begin{equation}\label{eq:pseudocounts}
\tilde{y}_i = \frac{p}{p + m}\frac{m_i}{2} + \frac{m}{p + m} y_i, \qquad (i=1,\dots,n).
\end{equation}
Each pseudo-count $\tilde{y}_i \in (0, m_i)$ is a convex combinations of the actual data and $m_i / 2$. This is equivalent to adding $ (p m_i) / (2m)$ successes and $(p m_i) / m$ trials to each $y_i$ and $m_i$, respectively, therefore shrinking the success proportions towards equiprobability and regularizing the estimation procedure. Importantly, it holds that $\tilde{\ell}(\beta; y) = (p / m + 1)\ell(\beta; \tilde{y})$ and therefore well-established algorithms for logistic regression may be used to maximize~\eqref{eq:penalized} even in presence of large datasets. In Section~\ref{sec:bayes}~and~\ref{sec:mle} we will show that the maximizer of~\eqref{eq:penalized} always exists and is unique. 
In addition, the considered conjugate prior penalty is a log-concave distribution, whose moments are always finite \citep{Diaconis1979,Chen2003}.

Adding small corrections to the response of a binomial model is a common regularization strategy. In the simplest case $p = n = 1$, correction~\eqref{eq:pseudocounts} reduces to the familiar bias-reducing form of the empirical logit \citep{Haldane1955, Anscombe1956}. In more general settings, these penalties have been referred to as \emph{data augmentation} priors \citep[e.g.,][]{Greenland2015}. The scheme of \citet{Clogg1991} is closely related to~\eqref{eq:pseudocounts}, albeit with several critical distinctions. Their correction amounts to adding $ (p \sum_{i=1}^n y_i) / (n m)$ successes and $p/n$ trials to each $y_i$ and $m_i$, respectively. Thus, \citet{Clogg1991} approach shrinks the success proportions towards the mean $\sum_{i=1}^n y_i/ m$ rather than $1/2$, which is a key aspect if one aims at reducing the bias \citep{Cordeiro1991, Firth1993}. Furthermore, the correction of \citet{Clogg1991} depends on the specific aggregation of the data and it leads to a different amount of shrinkage compared to \eqref{eq:pseudocounts}.

\section{Conjugate Bayes for logistic regression}\label{sec:bayes}
\subsection{Diaconis and Ylvisaker conjugate priors}
\label{sec:dy}

Bayesian inference is based on the posterior law $ p(\beta \mid y) = C(y) p(\beta)\exp\{\ell(\beta ; y)\}$, where $C(y)$ is the normalizing constant. Let $\tau > 0$ and let $\beta_0 \in \mathds{R}^p$ be a vector of hyperparameters. Moreover, let us define a vector of real numbers $\kappa = (\kappa_1,\dots,\kappa_n)^\T$ such that $\kappa_i  = m_i \exp(x_i^\T\beta_0) / \{1 + \exp(x_i^\T\beta_0)\}\in (0, m_i)$ for $i=1,\dots,n$. Thus, the conjugate prior of \citet{Diaconis1979} for a logistic regression model is
\begin{equation}\label{eq:prior}
p(\beta) = C \exp\left[\tau \sum_{i=1}^n\kappa_i (x_i^\T\beta) - \tau \sum_{i=1}^n m_i \log\{1 + \exp(x_i^\T\beta)\}\right],
\end{equation}
where the normalizing constant $0 < C < \infty$, albeit generally not available in closed form, is necessarily finite \citep[Theorem 1,][]{Diaconis1979}; additional considerations about~\eqref{eq:prior} can be found in \citet{Chen2003, Greenland2003}. The location parameter $\beta_0$ is the mode of the prior distribution and hence each ratio $\kappa_i / m_i \in (0,1)$  can be interpreted as the prior guess for the success probability $\pi_i$. Instead, the parameter $\tau$ controls the variability and quantifies the strength of our prior beliefs about $\beta_0$. More precisely, when $\tau \rightarrow 0$ then~\eqref{eq:prior} reduces to a uniform improper prior for $\beta$, whereas as $\tau \rightarrow \infty$ the prior converges to a point mass at $\beta_0$. The choice $\tau = 1$ places equal weight to the prior and the likelihood, therefore one typically focuses on the case $\tau \in (0,1)$. In the special case of binomial model with $p = n = 1$, the prior \eqref{eq:prior} induces the usual conjugate prior on the probability $\pi = \exp(\beta)/ \{ 1 + \exp(\beta)\}$, namely $\pi \sim \text{Beta}\{\tau\sum_{i=1}^n \kappa_i, \tau (m - \sum_{i=1}^n \kappa_i) \}$, with $m = \sum_{i=1}^n m_i$. 

Application of Bayes theorem to the binomial log-likelihood~\eqref{eq:loglik} under the prior~\eqref{eq:prior} leads to the following posterior distribution
\begin{equation}\label{eq:posterior}
p(\beta \mid y) = C(y) \exp\left[(\tau + 1)\sum_{i=1}^ny^*_i (x_i^\T\beta) - (\tau + 1) \sum_{i=1}^n m_i \log\{1 + \exp(x_i^\T\beta)\}\right],
\end{equation}
where $C(y) > 0$ is the normalizing constant and $y^* = (y^*_1,\dots,y^*_n)^\T$ is a vector of pseudo-counts such that $y^*_i = \kappa_i \tau / (\tau + 1) + y_i / (\tau + 1) \in (0, m_i)$, for $i=1,\dots,n$. The posterior is still in the class of Diaconis and Ylvisaker distributions, with updated location $y^*$ and precision $\tau + 1$. 
As we shall discuss in Section~\ref{sec:mle}, in light of the connection with \citet{Firth1993}, we propose a natural default choice with $\beta_0 = (0,\dots,0)^\T$ and $\tau = p/ m $, implying that $\kappa_i = m_i / 2$, for $i=1,\dots,n$. 

\subsection{Posterior inference}

The posterior law~\eqref{eq:posterior} can be expressed as $p(\beta \mid y) = C(y) \exp\{(\tau + 1) \ell(\beta; y^*)\}$,
that is, the posterior distribution is a function of the log-likelihood evaluated on the set of pseudo-counts $y^*$. Hence, the maximum a posteriori coincides with the maximizer of the log-likelihood~$\ell(\beta; y^*)$. Heuristically, the pseudo-counts regularize the estimation problem, as each $y_i^*$ belongs to the open set $(0, m_i)$. Consequently, the mode of the posterior distribution always exists, effectively solving the separability issue. This is formalized in the following
Theorem~\ref{teo1}, which refers to a general set of parameters $\beta_0$ and $\tau$, but it clearly applies also to the penalized log-likelihood~\eqref{eq:penalized}, that is when $\beta_0 = (0,\dots,0)^\T$ and $\tau = p/m$; refer to the Supplementary Materials for a proof, which is a natural consequence of results by \citet{Diaconis1979} and well-known properties of exponential families.

\begin{theorem}\label{teo1} Let $X$ be of full rank. Then the  mode of the posterior distribution~\eqref{eq:posterior}, corresponding to the maximizer of the  penalized likelihood
\begin{equation*}
\ell^*(\beta; y) =  (\tau + 1 ) \ell(\beta; y^*) = \ell(\beta; y) + \tau \sum_{i=1}^n\kappa_i (x_i^\T\beta) - \tau \sum_{i=1}^n m_i \log\{1 + \exp(x_i^\T\beta)\}
\end{equation*}
 with respect to $\beta$, exists and is unique. 
\end{theorem}

The mode of the posterior distribution~\eqref{eq:posterior} can be also regarded as the minimizer of the posterior expectation of a suitable entropy loss function \citep{Robert1996}. Hence, the mode is a formal Bayesian estimator with strong theoretical foundations. Bayesian estimators obtained under the entropy loss are invariant under reparametrizations. To clarify, if $\hat{\beta}_{\textsc{dy}}$ denotes the maximizer of~\eqref{eq:posterior}, then $\exp(\hat{\beta}_\textsc{dy})$ is the Bayesian estimator of the odds-ratios under the entropy loss.

The optimal value $\hat{\beta}_\textsc{dy}$ maximizing $\ell^*(\beta; y) =  (\tau + 1 ) \ell(\beta; y^*)$ can be found through standard Fisher scoring or expectation-maximization \citep{Durante2019} algorithms for logistic regression, considering the pseudo-counts $y^*$ in place of the original binomial data~$y$. 
Alternative algorithms could be also exploited, including quasi-Newton or conjugate gradient ascent methods \citep[e.g.,][]{Nocedal2006}. In all settings, the existence and uniqueness of $\hat{\beta}_\textsc{dy}$ are guaranteed by Theorem~\ref{teo1}, whereas the concavity of $\ell^*(\beta; y)$ implies that if an algorithm converges to a stationary point, then it must be the global optimum. Finally, full Bayesian inference based on the posterior~\eqref{eq:posterior} can be performed via Markov chain Monte Carlo. For instance, the data-augmentation scheme of \citet{Polson2013} may be easily adapted to this setting, leading to a simple Gibbs sampling algorithm. % 
\section{Penalized maximum likelihood}\label{sec:mle}
\subsection{Shrinkage, bias reduction and data aggregation}

The maximum likelihood estimates of $\beta$ in a logistic regression model are biased away from the point $\beta = 0$; refer for example to \citet{McCullagh1989, Cordeiro1991}. Thus, bias correction requires a certain amount of shrinkage towards that point. In \citet{Firth1993} this is achieved through a Jeffrey's prior penalty, whose mode is indeed equal to $0$ \citep{Kosmidis2021}. In a similar fashion, we set the mode $\beta_0 = (0,\dots,0)^\T$ in the prior specification~\eqref{eq:prior}, which implies that $\kappa_i = m_i / 2$, for $i=1,\dots,n$. The amount of shrinkage is regulated by the precision parameter $\tau$, whose choice is more delicate. Heuristically, one may set $\tau$ proportional to $m^{-1}$, so that for $m$ large enough the contribution of the prior becomes negligible compared to the weight of the likelihood in~\eqref{eq:posterior}. In particular, if $\tau = p/m$ then the amount of shrinkage is proportional to the model complexity, which seems desirable. These choices lead to the penalized log-likelihood \eqref{eq:penalized} and the pseudo-counts~\eqref{eq:pseudocounts}. 
Such an intuitive choice ensures an accurate approximation of the reduced-bias estimators of \citet{Firth1993}, and that the resulting prior distribution is invariant under different aggregation of the data, in contrast with \citet{Clogg1991}.

\subsection{Penalized score equations and connection with Firth (1993)}
\label{sec:firth_connection}

Firth's penalized maximum likelihood estimate is obtained as the solution of the system of penalized score equations $U_{r, \textsc{fi}}(\beta) = 0$ for $r=1,\dots,p$, having defined
\begin{equation*}
U_{r, \textsc{fi}}(\beta) = \sum_{i=1}^n (y_i - m_i \pi_i)x_{ir} - p \sum_{i=1}^n \left(\frac{h_i}{p}\right) (\pi_i - 1/2)x_{ir}, \quad (r=1,\dots,p),
\end{equation*}
where $h_1,\dots,h_n$ represent the diagonal elements of the $n \times n$ projection matrix 
$$
H(\beta) = W(\beta)^{1/2}X\{X^\T W(\beta) X\}^{-1}X^\T W(\beta)^{1/2},
$$ 
with $W(\beta) = \text{diag}\{m_1\pi_1 (1-\pi_1),\dots, m_n \pi_n (1-\pi_n)\}$. 
Instead, the maximizer of the penalized log-likelihood~\eqref{eq:penalized} corresponds to the solution of the system of penalized score equations $U_{r}(\beta) = 0$ for $r=1,\dots,p$, where
\begin{equation*}
U_r(\beta) = \frac{\partial}{\partial \beta_r}\tilde{\ell}(\beta; y) = \sum_{i=1}^n (y_i - m_i \pi_i)x_{ir} - p\sum_{i=1}^n \left(\frac{m_i}{m}\right) (\pi_i - 1/2)x_{ir}, \quad (r=1,\dots,p).
\end{equation*}
Clearly, the penalized scores $U_{r, \textsc{fi}}(\beta)$ and $U_r(\beta)$ differs only in their penalty term. Moreover, the matrix $H(\beta)$ has rank $p$ and is idempotent and symmetric, implying that the sum of its diagonal terms is $\sum_{i=1}^n h_i = p$. Hence, the following approximation holds
\begin{equation}\label{eq:approx}
\sum_{i=1}^n \left(\frac{h_i}{p}\right) (\pi_i - 1/2)x_{ir}\approx  \sum_{i=1}^n \left(\frac{m_i}{m}\right) (\pi_i - 1/2)x_{ir}, \quad (r=1,\dots,p),
\end{equation}
Our method considers the mean of the values $(\pi_i - 1/2)x_{ir}$, with weights $m_i / m$, instead of the mean of the same quantities but with weights $h_i/ p$, as in \citet{Firth1993}. Hence, the penalized score functions $U_{r, \textsc{fi}}(\beta)$ and $U_r(\beta)$ are approximately interchangeable, as well as the corresponding solutions. 
From a computational perspective, Firth's modified score equations can be solved via quasi-Fisher scoring \citep[e.g.][]{Kosmidis2010}. However, this approach can be computationally challenging in settings with large $n$ and $p$ since the terms $h_1, \dots, h_n$ from $H(\beta)$ depend on the current value of $\beta$. The penalized scores, instead, can be solved with any software for logistic likelihood optimization, and therefore provide  a computationally convenient option to approximate Firth's scores in large dimensions.

To shed further light on the approximation~\eqref{eq:approx}, let us consider a disaggregated representation of the data, in which the response variables $y_{ij} \in\{0,1\}$ are binary indicators, for $j=1,\dots,m_i$ and $i=1,\dots,n$, so that $y_i = \sum_{j=1}^{m_i}y_{ij}$. Each $y_{ij}$ is associated with the covariate vector $x_i$, which is repeated $m_i$ times in the disaggregated design matrix. Furthermore, let $h_{ij}$ be the leverage values of such a binary regression models, for $j=1,\dots,m_i$ and $i=1,\dots,n$. We can rewrite the penalties appearing in~\eqref{eq:approx} in terms of these disaggregated quantities, obtaining that $p^{-1} \sum_{i=1}^n h_i (\pi_i - 1/2)x_{ir} = p^{-1} \sum_{i=1}^n\sum_{j=1}^{m_i} h_{ij}(\pi_i - 1/2)x_{ir}$ and  $m^{-1} \sum_{i=1}^n m_i (\pi_i - 1/2)x_{ir} = m^{-1}\sum_{i=1}^n\sum_{j=1}^{m_i} (\pi_i - 1/2)x_{ir}$. The first equality is a consequence of the invariance property of \citet{Firth1993} under alternative representations of the data. Hence, approximation~\eqref{eq:approx} can be equivalently written as follows
\begin{equation}\label{eq:approx2}
\sum_{i=1}^n\sum_{j=1}^{m_i} \left(\frac{h_{ij}}{p}\right)(\pi_i - 1/2)x_{ir}\approx \frac{1}{m}\sum_{i=1}^n\sum_{j=1}^{m_i} (\pi_i - 1/2)x_{ir}, \quad (r=1,\dots,p),
\end{equation}
where the approximation is due to the replacement of a weighted mean with a simple arithmetic mean. Importantly, we have that  $m^{-1} \sum_{i=1}^n\sum_{j=1}^{m_i}h_{ij} = p/m$, clarifying that equations~\eqref{eq:approx} and \eqref{eq:approx2} rely on the substitution of the leverages $h_{ij}$ with their mean $p/m$. The condition $h_{ij} \approx p / m$ is known as \emph{approximate quadratic balance} and has been exploited by \citet{Cordeiro1991} to obtain the reduced bias estimate $(1 - p/m)\hat{\beta}$, which indeed deflates the maximum likelihood estimate towards zero by the factor $1 - p / m$. The quadratic balance condition holds exactly true in some special cases. For instance, when $p = n = 1$ then both \citet{Firth1993} and our approach correspond to a $\text{Beta}(1/2, 1/2)$ prior penalty for $\pi$ and coincide with the empirical logit correction of \citet{Haldane1955}. More generally, in any saturated model with $p = n$ and with balanced number of trials $m_1 = \dots = m_n$ the two approaches formally agree, since we must have that $\sum_{i=1}^n h_i = n$ and $0 \le h_i \le 1$, implying that $h_i = 1$. Replacing the diagonal elements of a projection matrix with their mean also appears in seemingly unrelated settings, for example to obtain the so-called generalized cross-validation index \citep[][§5.3]{Wasserman2005}.

\subsection{Asymptotic properties and inference}

Consistency of the penalized maximum likelihood estimator $\hat{\beta}_\textsc{dy}$ follows directly from the consistency of the maximum likelihood estimator $\hat{\beta}$. Moreover, $\hat{\beta}$ and $\hat{\beta}_{\textsc{dy}}$ have the same asymptotic distribution and hence are both asymptotically unbiased and efficient. Indeed, the penalty terms described in~\eqref{eq:approx} are of order $O(1)$ and hence dominated by the score function, as discussed in \citet{Firth1993, Pagui2017}. Thus, $\hat{\beta}$ and $\hat{\beta}_{\textsc{dy}}$ are both asymptotically distributed as multivariate Gaussians centered on the true value $\beta^\dag$ and with asymptotic inverse covariance matrix $I(\beta^\dag) = X^\T W( \beta^\dag) X$. Consequently, Wald-type confidence intervals may be constructed by plugging-in $\hat{\beta}_\textsc{dy}$ into $I(\beta)$ and then proceeding in the usual manner.

\section{Illustrations}

\subsection{Infant birthweight study}\label{sec:birth}
We replicate an example presented in \citet{Kosmidis2020}, which considers a study of low birthweight.
Data comprises $n = 100$ births and the binary outcome of interest is a dichotomization of infant birthweight (below or above $2.5$ kilograms). The probability of low birthweight is modelled as a function of an intercept and six covariates about the mother.
The maximum likelihood estimate $\hat{\beta}$ of the regression coefficients $\beta = (\beta_1,\dots,\beta_7)^\T$ exists and is finite.
We simulate $10000$ datasets from a logistic regression model with parameter~$\hat{\beta}$ and evaluate the inferential properties of the proposed estimator, comparing them with popular bias-correction methods in a regular scenario. 

\begin{table}[!tb]
\begin{tabular}{llrrrrrrr}
\toprule
& & $\beta_1$ & $\beta_2$ & $\beta_3$ & $\beta_4$ & $\beta_5$ & $\beta_6$ & $\beta_7$ \\
\midrule
Bias & Maximum likelihood $\hat{\beta}$                       & -1.42 & -0.01 & 0.09  & -0.04 & -0.18 & -0.12 & 0.34 \\
     & Penalized maximum likelihood $\hat{\beta}_\textsc{dy}$ & -0.08 & 0.00  & -0.01 & 0.03  & -0.01 & 0.03  & 0.00 \\
     & \citet{Clogg1991}                                      & -0.22 & 0.00 & 0.00  & 0.02  & -0.01 & 0.03  & 0.05 \\
     & \citet{Firth1993}                                      & -0.08 & 0.00 & 0.01  & 0.00 & 0.00 & 0.00 & 0.02 \\
     & \citet{Pagui2017}                                      & -0.38 & 0.00 & 0.03  & -0.01 & -0.06 & -0.03 & 0.10 \\
\midrule
\textsc{rmse} & Maximum likelihood $\hat{\beta}$                       & 6.88 & 0.06 & 0.64 & 0.65 & 0.81 & 1.12 & 1.49 \\
              & Penalized maximum likelihood $\hat{\beta}_\textsc{dy}$ & 5.71 & 0.05 & 0.54 & 0.56 & 0.71 & 0.96 & 1.22 \\
              & \citet{Clogg1991}                                      & 5.83 & 0.05 & 0.55 & 0.57 & 0.70 & 0.97 & 1.25 \\
              & \citet{Firth1993}                                      & 5.94 & 0.05 & 0.57 & 0.58 & 0.71 & 0.95 & 1.28 \\
              & \citet{Pagui2017}                                      & 6.12 & 0.06 & 0.58 & 0.60 & 0.78 & 1.01 & 1.31 \\
\bottomrule
\end{tabular}
\caption{Simulation study on the low birthweight study. \textsc{rmse}: root mean squared error. \label{tab2}}

\end{table}

Results are presented in  Table~\ref{tab2}. The values for the maximum likelihood are computed excluding samples with data separation, occurring in $100$ replications out of $10000$.
The maximum likelihood estimator performs significantly worse than all the reduced-bias methodologies in terms of bias and root mean squared error.
Moreover, we observe a striking empirical similarity in terms of bias between the performance of the proposed penalized estimator $\hat{\beta}_\textsc{dy}$ and the one of \citet{Firth1993}; this finding is in line with the considerations discussed in Section~\ref{sec:firth_connection}. Empirical results also confirm a limitation of \citet{Clogg1991}, which fails at reducing the bias for the intercept parameter $\beta_1$, since it shrinks the response towards the empirical proportion of successes rather than $1/2$. Finally, the approach of \citet{Pagui2017} is designed for correcting median bias and therefore is expected to be worse, in terms of bias, than \citet{Firth1993}. The proposed penalized estimator $\hat{\beta}_\textsc{dy}$ slightly improves the mean squared error compared to \citet{Firth1993}. This aspect seems to be empirically confirmed also in the following illustrative example.

\subsection{High dimensional synthetic dataset}

\label{sec:sim_hd}
\begin{figure}
\includegraphics[width = \textwidth]{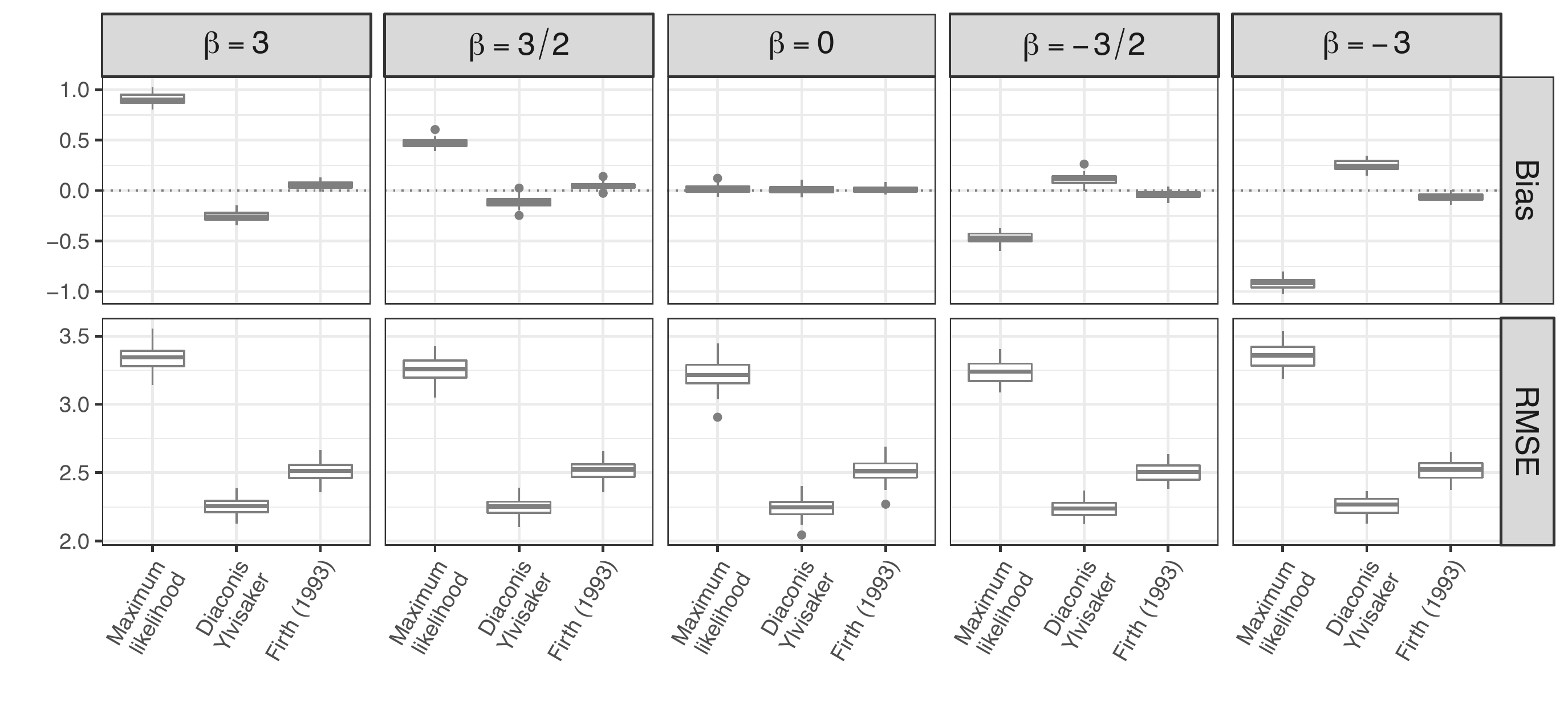} % theme_bw()
\caption{Bias and root mean squared error (\textsc{rmse}) of maximum likelihood and reduced-bias estimates, for the parameters of a logistic regression model. Boxplots indicate variability across groups of regression coefficients.}
\label{fig2}
\end{figure}

We finally consider a synthetic dataset that mimicks the high dimensional scenario described in \citet{Sur2019}. In particular, we simulate $5000$ datasets from a binary logistic regression model with $n = 1000$ observations and $p = 200$. Covariates are sampled from a normal distribution with mean $0$ and variance $1/n$. Moreover, regression coefficients are divided in $5$ blocks of size $40$, whose values are $\{-3,-3/2,0,3/2,3\}$.  The purpose of this study is to investigate the performance of the proposed estimator in computationally challenging scenarios. 

In first place, we shall note that the execution times of the estimation procedure for $\hat{\beta}_\textsc{dy}$, obtained via scalable algorithms for logistic regression, are orders of magnitude faster than the \texttt{brglm2} implementation of \citet{Firth1993}. For instance, a single replication with $n =1000$ and $p=200$ required an average elapsed time of $1.24$ milliseconds in the former case, against approximately half a second in the latter, on a 2020 Macbook Pro. These differences are even more marked for larger values of $n$ and $p$, to the extent that \citet{Firth1993} estimates could not be obtained within hours of running time with $n=10000$, $p=2000$ and correlated design matrix $X$.

Secondly, we computed the bias and the root mean squared error of maximum likelihood and reduced-bias estimators, which are depicted in Figure~\ref{fig2}. Current empirical findings confirm the poor behaviour of maximum likelihood estimates, consistently with the existing literature \citep{Kosmidis2021}. The proposed correction provides an important improvement in terms of bias reduction compared to the maximum likelihood estimator, although the bias is slightly larger than that of \citet{Firth1993}. This is expected and consistent with the findings of Sections~\ref{sec:firth_connection} and \ref{sec:birth}. Furthermore, our penalized procedure achieves a lower mean squared error compared to both the maximum likelihood and the approach of \citet{Firth1993}. This does not come as a contradiction, since \citet{Firth1993} approach does not explicitly reduce the mean squared error. 
% This empirical finding is likely due to the different tail behaviour of \citet{Diaconis1979} priors compared to Jeffrey's priors, the latter being more dispersed and displaying heavier tails, as discussed in Section~\ref{sec:dy}.  Similar qualitative considerations can be also found in \citet{Greenland2015}. 
This empirical finding is likely due to the different tail behavior of \citet{Diaconis1979} priors compared to Jeffrey's priors, the latter being more dispersed and displaying heavier tails.

\section*{Acknowledgements} 
The authors wish to thank David Dunson, Sonia Migliorati and Aldo Solari for their insightful comments on an early draft of this manuscript.

\appendix
\section*{Supplementary material}
\label{SM}
 Supplementary material includes a detailed proof of Theorem~\ref{teo1}, a graphical illustration of the proposed prior and an illustrative case study with separability issues. Software is available online at \url{github.com/tommasorigon/logistic-bias-reduction}.

\section{Proof of Theorem 1}
The posterior mode of $p(\beta \mid y)$ coincides with the optimizer of the quantity $\ell(\beta; y^*)$, namely the log-likelihood evaluated on the vector of pseudo-counts $y^*$.  The Fisher information matrix is
\begin{equation*}
I(\beta) := - \frac{\partial^2}{\partial \beta \partial \beta^\T}\ell(\beta; y^*) =  X^\T W(\beta)X, \:   W(\beta) = \text{diag}\{m_1 \pi_1 (1-\pi_1),\dots, m_n \pi_n (1-\pi_n)\},
\end{equation*}
which does not depend on $y^*$. Since $X$ is full rank, then the matrix $I(\beta)$ is positive definite for any value of $\beta$, a well known result. Hence, the optimal value $\hat{\beta}$ maximizing $\ell(\beta; y^*)$, if it exists, is unique. Since $X$ is full rank, then logistic regression model is a regular $p$-dimensional exponential family and therefore the optimal value of $\ell(\beta; y^*)$ exists if and only if the sufficient statistic 
\begin{equation*}
s = (s_1,\dots,s_p)^\T = \left(\sum_{i=1}^nx_{i1} y^*_i, \dots, \sum_{i=1}^n x_{ip}y^*_i\right)^\T,
\end{equation*}
belongs to the interior of the closed convex hull of the support of $s$; see Theorem~5.8 in \citet{Pace1997}. This is indeed the case, as a consequence of Theorem~1 in \citet{Diaconis1979}. % Note that this is not always the case for general vectors of counts $y_1,\dots,y_n$. 

\section{Graphical comparison}
\begin{figure}[tb]
\includegraphics[width= \textwidth]{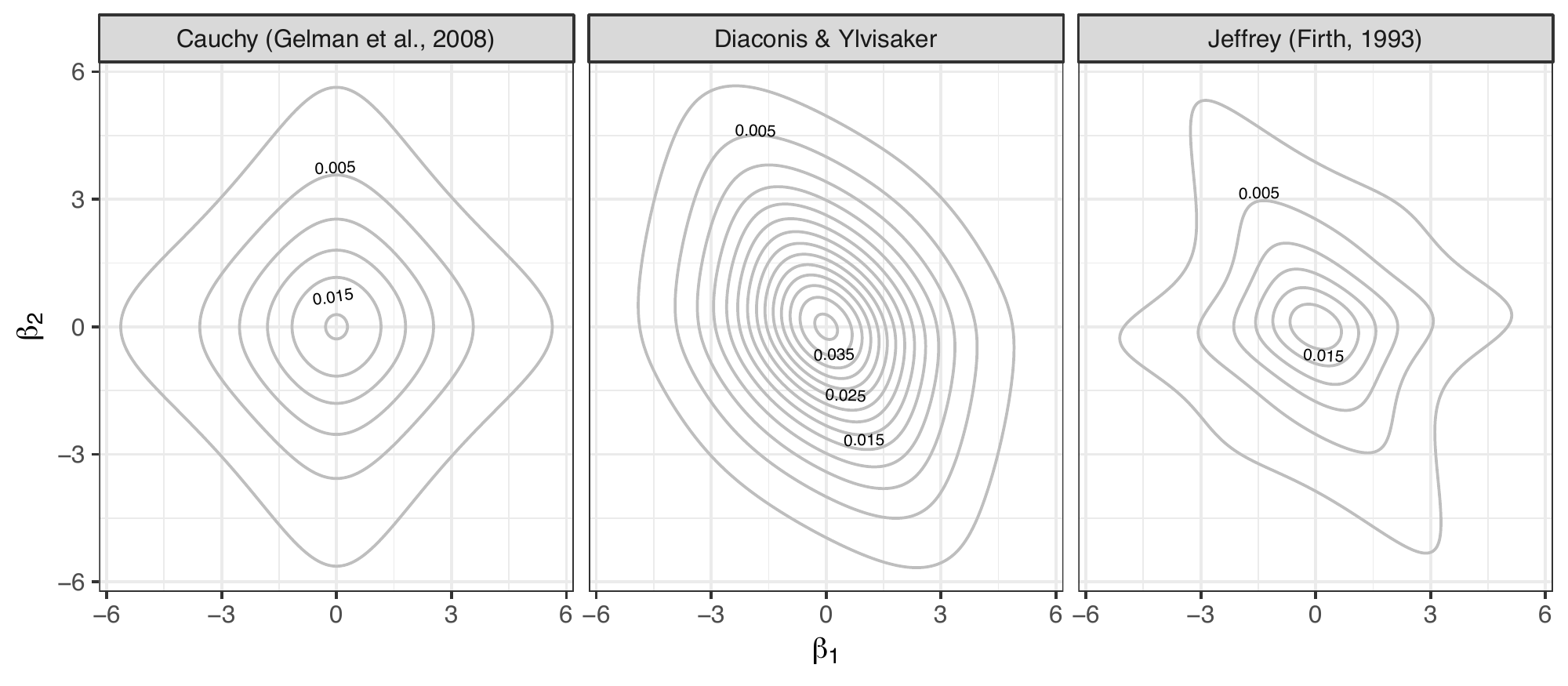}
\caption{Contour levels of different priors for the coefficients $\beta = (\beta_1,\beta_2)^\T$ of a logistic regression model with no intercept. Covariates $x_1,\dots,x_n$ have been standardized, with $n = 10$ and $p=2$. From left to right: Cauchy prior of \citet{Gelman2008}, conjugate prior of  \citet{Diaconis1979} with $\beta_0 = (0,\dots,0)^\T$ and $\tau = p/ m $, and Jeffrey's prior of \citet{Firth1993}. }
\label{fig1}
\end{figure}

We provide further insights on the proposed Diaconis-Ylvisaker prior through a graphical comparison with the Cauchy prior of \citet{Gelman2008}, and the Jeffrey's prior of \citet{Firth1993}. 
The Cauchy prior has been suggested as a default choice for  regularizing logistic regression, and is a routinely used in statistical software as a default specification.
In this illustrative example, we focus on $p=2$ standardized covariates with $n = 10$ observations.
Figure~\ref{fig1} depicts the contour plots of the different priors.
The locations and scales of quite similar, with the distibutions centered on $(0,0)$ and assigning most of their mass in the interval $[-5,5] \times [-5,5]$.
Instead, their shapes are markedly different. 
In particular, the Diaconis-Ylvisaker and Jeffrey's prior depend on the specific covariates values and reflect correlations in the predictors.
In addition, the  \citet{Diaconis1979} prior is roughly ellipsoidal and with lighter tails compared to Cauchy and Jeffrey priors.

\section{Endometrial cancer study}\label{sec:endo}

\begin{table}[b]
\begin{tabular}{lrrrr}
\toprule
 & $\beta_1$ & $\beta_2$ & $\beta_3$ & $\beta_4$ \\
\midrule
Maximum likelihood $\hat{\beta}$ & 4.305 (1.637) & $+\infty$ ($+\infty$) & -0.042 (0.044) & -2.903 (0.846)\\
Penalized maximum likelihood $\hat{\beta}_\textsc{dy}$ & 3.579 (1.459) & 3.431 (1.893) & -0.034 (0.040) & -2.458 (0.748)\\
\citet{Clogg1991} &3.622 (1.471) & 3.223 (1.722) & -0.034 (0.040) & -2.511 (0.761)  \\
\citet{Firth1993} & 3.775 (1.489) & 2.929 (1.551) & -0.035 (0.040) & -2.604 (0.776)\\
\citet{Pagui2017} & 3.969 (1.552) & 3.869 (2.298) & -0.039 (0.042) & -2.708 (0.803) \\
\bottomrule
\end{tabular}
\caption{Estimated regression coefficients on the endometrial cancer study (with standard errors in parentheses)\label{tab1}}
\end{table}

% The practical benefits of the proposed penalized estimates are illustrated through examples of increasing complexity. 
% As a first illustration, we consider the endometrial cancer grade dataset \citep{Heinze2002}, a study on $n = 79$ patients which aims at evaluating the relationship between the histology of the endometrium (low against high), and three risk factors.
We consider the endometrial cancer grade dataset \citep{Heinze2002}, a study on $n = 79$ patients which aims at evaluating the relationship between the histology of the endometrium (low against high), and three risk factors.
A logistic regression model was fitted with parameter vector $\beta=(\beta_1, \beta_2,\beta_3,\beta_4)^\T$, with the first coefficient corresponding to an intercept term and the remaining to neovasculation, pulsatility index of arteria uterina and endometrium height, respectively. 
As shown in \citet{Heinze2002}, the maximum likelihood estimate does not exist since the estimated value for the coefficient $\beta_2$ associated with neovasculation is divergent due to quasi-separability. The omission of the neovasculation information from the set of covariates is entirely inappropriate, as the other factors would not be properly adjusted for this highly informative risk factor; refer to \citet{Kosmidis2021} for a discussion on the implications of separability on inferential procedures.
We therefore compare the proposed penalized maximum likelihood estimate $\hat{\beta}_\textsc{dy}$ with the approaches of \citet{Clogg1991} and \citet{Firth1993}. We also consider the median unbiased estimators of \citet{Pagui2017}, \citet{Kosmidis2020}. Computations are performed using the \texttt{brglm2} R package and the R function \texttt{glm}.

Estimates for the regression coefficients $\beta = (\beta_1, \beta_2,\beta_3,\beta_4)^\T$ and the corresponding standard errors are reported in Table~\ref{tab1}. As expected, all the reducing-bias methods produce finite estimates for $\beta_2$, which are all close to $3$ and indicate a strong effect of neovasculation on the response probability. Point estimates and standard errors for the remaining coefficients are also quite similar. The approach of \citet{Clogg1991} and our penalized method lead to similar estimates since the former shrinks the estimated probabilities towards the sample proportion, which in this case is $\sum_{i=1}^n y_i / m = 0.38$. 
Despite the stated aim of \citet{Clogg1991} was not performing bias reduction, it performs reasonably well in this example, and the reasons for its good empirical performance are likely due to its similarity with our approach, which in turn approximates the one of \citet{Firth1993}.

\bibliographystyle{chicago}
\bibliography{biblio}

\begin{thebibliography}{}

\bibitem[\protect\citeauthoryear{Albert and Anderson}{Albert and
  Anderson}{1984}]{Albert1984}
Albert, A. and J.~A. Anderson (1984).
\newblock {On the existence of maximum likelihood estimates in logistic
  regression models}.
\newblock {\em Biometrika\/}~{\em 71\/}(1), 1--10.

\bibitem[\protect\citeauthoryear{Anscombe}{Anscombe}{1956}]{Anscombe1956}
Anscombe, F.~G. (1956).
\newblock On estimating binomial response relations.
\newblock {\em Biometrika\/}~{\em 43}, 461--464.

\bibitem[\protect\citeauthoryear{Chen and Ibrahim}{Chen and
  Ibrahim}{2003}]{Chen2003}
Chen, M.~H. and J.~G. Ibrahim (2003).
\newblock {Conjugate priors for generalized linear models}.
\newblock {\em Statist. Sin.\/}~{\em 13\/}(2), 461--476.

\bibitem[\protect\citeauthoryear{Clogg, Rubin, Schenker, Schultz, and
  Weidman}{Clogg et~al.}{1991}]{Clogg1991}
Clogg, C.~C., D.~B. Rubin, N.~Schenker, B.~Schultz, and L.~Weidman (1991).
\newblock {Multiple imputation of industry and occupation codes in census
  public-use samples using Bayesian logistic regression}.
\newblock {\em J. Am. Statist. Assoc.\/}~{\em 86\/}(413), 68--78.

\bibitem[\protect\citeauthoryear{Cordeiro and McCullagh}{Cordeiro and
  McCullagh}{1991}]{Cordeiro1991}
Cordeiro, G.~M. and P.~McCullagh (1991).
\newblock {Bias Correction in Generalized Linear Models}.
\newblock {\em J. R. Statist. Soc. B\/}~{\em 53\/}(3), 629--643.

\bibitem[\protect\citeauthoryear{Diaconis and Ylvisaker}{Diaconis and
  Ylvisaker}{1979}]{Diaconis1979}
Diaconis, P. and D.~Ylvisaker (1979).
\newblock {Conjugate prior for exponential families}.
\newblock {\em Ann. Statist.\/}~{\em 7\/}(2), 269--292.

\bibitem[\protect\citeauthoryear{Durante and Rigon}{Durante and
  Rigon}{2019}]{Durante2019}
Durante, D. and T.~Rigon (2019).
\newblock {Conditionally conjugate mean-field variational Bayes for logistic
  models}.
\newblock {\em Statist. Sc.\/}~{\em 34\/}(3), 472--485.

\bibitem[\protect\citeauthoryear{Firth}{Firth}{1993}]{Firth1993}
Firth, D. (1993).
\newblock {Bias reduction of maximum likelihood estimates}.
\newblock {\em Biometrika\/}~{\em 80\/}(1), 27--38.

\bibitem[\protect\citeauthoryear{Gelman, Jakulin, Pittau, and Su}{Gelman
  et~al.}{2008}]{Gelman2008}
Gelman, A., A.~Jakulin, M.~G. Pittau, and Y.~S. Su (2008).
\newblock {A weakly informative default prior distribution for logistic and
  other regression models}.
\newblock {\em Ann. Appl. Statist.\/}~{\em 2\/}(4), 1360--1383.

\bibitem[\protect\citeauthoryear{Greenland}{Greenland}{2003}]{Greenland2003}
Greenland, S. (2003).
\newblock {Generalized conjugate priors for Bayesian analysis of risk and
  survival regressions}.
\newblock {\em Biometrics\/}~{\em 59\/}(1), 92--99.

\bibitem[\protect\citeauthoryear{Greenland and Mansournia}{Greenland and
  Mansournia}{2015}]{Greenland2015}
Greenland, S. and M.~A. Mansournia (2015).
\newblock {Penalization, bias reduction, and default priors in logistic and
  related categorical and survival regressions}.
\newblock {\em Statist. Med.\/}~{\em 34\/}(23), 3133--3143.

\bibitem[\protect\citeauthoryear{Haldane}{Haldane}{1955}]{Haldane1955}
Haldane, J. S.~B. (1955).
\newblock The estimation and significane of the logarithm oa ration of
  frequencies.
\newblock {\em Ann. Hum. Gen.\/}~{\em 20}, 309--311.

\bibitem[\protect\citeauthoryear{Heinze and Schemper}{Heinze and
  Schemper}{2002}]{Heinze2002}
Heinze, G. and M.~Schemper (2002).
\newblock {A solution to the problem of separation in logistic regression}.
\newblock {\em Statist. Med.\/}~{\em 21\/}(16), 2409--2419.

\bibitem[\protect\citeauthoryear{{Kenne Pagui}, Salvan, and Sartori}{{Kenne
  Pagui} et~al.}{2017}]{Pagui2017}
{Kenne Pagui}, E.~C., A.~Salvan, and N.~Sartori (2017).
\newblock {Median bias reduction of maximum likelihood estimates}.
\newblock {\em Biometrika\/}~{\em 104\/}(4), 923--938.

\bibitem[\protect\citeauthoryear{Kosmidis and Firth}{Kosmidis and
  Firth}{2010}]{Kosmidis2010}
Kosmidis, I. and D.~Firth (2010).
\newblock {A generic algorithm for reducing bias in parametric estimation}.
\newblock {\em Electr. J. Statist.\/}~{\em 4}, 1097--1112.

\bibitem[\protect\citeauthoryear{Kosmidis and Firth}{Kosmidis and
  Firth}{2021}]{Kosmidis2021}
Kosmidis, I. and D.~Firth (2021).
\newblock {Jeffreys-prior penalty, finiteness and shrinkage in
  binomial-response generalized linear models}.
\newblock {\em Biometrika\/}~{\em 108\/}(1), 71--82.

\bibitem[\protect\citeauthoryear{Kosmidis, {Kenne Pagui}, and Sartori}{Kosmidis
  et~al.}{2020}]{Kosmidis2020}
Kosmidis, I., E.~C. {Kenne Pagui}, and N.~Sartori (2020).
\newblock {Mean and median bias reduction in generalized linear models}.
\newblock {\em Statist. Comp.\/}~{\em 30\/}(1), 43--59.

\bibitem[\protect\citeauthoryear{McCullagh and Nelder}{McCullagh and
  Nelder}{1989}]{McCullagh1989}
McCullagh, P. and J.~A. Nelder (1989).
\newblock {\em {Generalized linear models}\/} (Second ed.).
\newblock Springer.

\bibitem[\protect\citeauthoryear{Nocedal and Wright}{Nocedal and
  Wright}{2006}]{Nocedal2006}
Nocedal, J. and S.~Wright (2006).
\newblock {\em Conjugate Gradient Methods}.
\newblock New York, NY: Springer New York.

\bibitem[\protect\citeauthoryear{Pace and Salvan}{Pace and
  Salvan}{1997}]{Pace1997}
Pace, L. and A.~Salvan (1997).
\newblock {\em Principles of Statistical Inference from a Neo-Fisherian
  Perspective}.
\newblock World Scientific.

\bibitem[\protect\citeauthoryear{Polson, Scott, and Windle}{Polson
  et~al.}{2013}]{Polson2013}
Polson, N.~G., J.~G. Scott, and J.~Windle (2013).
\newblock {Bayesian inference for logistic models using Polya-Gamma latent
  variables}.
\newblock {\em J. Am. Statist. Assoc.\/}~{\em 108\/}(504), 1--42.

\bibitem[\protect\citeauthoryear{Puhr, Heinze, Nold, Lusa, and
  Geroldinger}{Puhr et~al.}{2017}]{Puhr2017}
Puhr, R., G.~Heinze, M.~Nold, L.~Lusa, and A.~Geroldinger (2017).
\newblock {Firth's logistic regression with rare events: accurate effect
  estimates and predictions?}
\newblock {\em Statist. Med.\/}~{\em 36\/}(14), 2302--2317.

\bibitem[\protect\citeauthoryear{Robert}{Robert}{1996}]{Robert1996}
Robert, C.~P. (1996).
\newblock {Intrinsic losses}.
\newblock {\em Th. Dec.\/}~{\em 40\/}(2), 191--214.

\bibitem[\protect\citeauthoryear{Sur and Cand{\`e}s}{Sur and
  Cand{\`e}s}{2019}]{Sur2019}
Sur, P. and E.~J. Cand{\`e}s (2019).
\newblock A modern maximum-likelihood theory for high-dimensional logistic
  regression.
\newblock {\em Proc. Nat. Acad. Sci.\/}~{\em 116}, 14516--25.

\bibitem[\protect\citeauthoryear{Wasserman}{Wasserman}{2005}]{Wasserman2005}
Wasserman, L. (2005).
\newblock {\em {All of nonparametric statistics}}.
\newblock Springer.

\end{thebibliography}

\end{document}